\documentclass[12pt,letterpaper]{article}
\newcommand{\beq}{\begin{equation}}
\newcommand{\eeq}{\end{equation}}
\newcommand{\beqr}{\begin{displaymath}}
\newcommand{\eeqr}{\end{displaymath}}
\newcommand{\beqa}{\begin{eqnarray}}
\newcommand{\eeqa}{\end{eqnarray}}
\newcommand{\beqar}{\begin{eqnarray*}}
\newcommand{\eeqar}{\end{eqnarray*}}
\newcommand{\al}{\alpha}

\newcommand{\del}{\delta}
\newcommand{\D}{\Delta}

\renewcommand{\L}{\Lambda}

\newcommand{\cR}{{\cal R}}
\newcommand{\cT}{{\cal T}}
\newcommand{\ssc}{\scriptscriptstyle}
\newcommand{\eg}{{\it e.g.,\ }}
\newcommand{\ie}{{\it i.e.,\ }}

\newcommand{\reef}[1]{(\ref{#1})}

\newcommand{\non}{\nonumber}
\newcommand{\pf}{\partial}

\newcommand{\ti}[1]{\tilde{#1}}  

\newcommand{\df}{\textrm{d}}
\newcommand{\mt}[1]{\textrm{\tiny #1}}
\newcommand{\bk}[1]{#1}
\newcommand{\br}[1]{{\cal #1}}
\newcommand{\inte}[1]{\ti{{\cal #1}}}
\newcommand{\GC}{G_{\ssc{D}}}
\newcommand{\w}{\omega}

\begin{document}

\thispagestyle{empty}
\rightline{\small hep--th/0106140 \hfill McGill/00-11}\nopagebreak
\vskip -.6ex
\vspace*{2cm}

\begin{center}
{\bf Consistency Conditions for Brane Worlds\\ in Arbitrary Dimensions}
\vspace*{1cm}

Fr\'ed\'eric Leblond,$^{}$\footnote{E-mail: fleblond@hep.physics.mcgill.ca}
Robert C. Myers$^{}$\footnote{E-mail: rcm@hep.physics.mcgill.ca} and 
David J. Winters$^{}$\footnote{E-mail: winters@hep.physics.mcgill.ca} 
\vspace*{0.3cm}

{\it Department of Physics, McGill University\\ 3600 University Street,
Montr\'eal, QC, H3A 2T8, Canada}\\
\vspace{2cm}
ABSTRACT
\end{center}
We consider ``brane world sum rules'' for compactifications involving
an arbitrary number of spacetime dimensions. One of the most striking results
derived from such consistency conditions is the necessity for negative tension
branes to appear in five--dimensional scenarios. We show how this result
is easily evaded for brane world models with more than five dimensions.
As an example,
we consider a novel realization of the Randall--Sundrum scenario in six
dimensions involving only positive tension branes.

\vfill
\setcounter{page}{0}
\setcounter{footnote}{0}
\newpage

\section{Introduction}

Brane world scenarios have captured the imagination of high
energy theorists because they provide new mechanisms 
for resolving many problems in particle physics which have long
resisted solution. In particular, this framework offers new explanations
for the small ratio between the symmetry breaking scale of electroweak
physics, and the Planck scale of quantum gravity. 
One possibility\cite{arkani,Anton} is that the observed four--dimensional
Planck scale is a derived quantity determined by the true fundamental scale,
which may be as low as 1 TeV, and the compactification scale,
which may be as large as a fraction of a millimeter. Another
alternative\cite{RS1,RS2} is that this small ratio of energy scales appears
as a gravitational redshift in a warped
compactification of a theory with a single fundamental scale.

In these scenarios with large extra dimensions, the distance scales involved
in compactification are much larger than that set by quantum gravity.
Therefore, in constructing such models, one must take seriously the question
of whether the spacetime geometry, along with the accompanying
background fields and branes, solves the classical Einstein
equations in higher dimensions. In ref.~\cite{gibbons}, the authors 
showed how, from Einstein's equations, one can derive a set of consistency
conditions or ``sum rules'' that must be satisfied by any such
model.\footnote{Similar analyses appeared earlier, in studying
warped compactifications without branes\cite{berd} and certain
five--dimensional brane world scenarios\cite{one}.}
One of their most striking results was to show that for a very broad class
of models, \eg  those given in refs.~\cite{RS1,wise,kallosh}, a consistent
compactification demands the inclusion of negative tension branes.
This is a rather disappointing conclusion as negative tension branes are 
inherently unstable objects, although these instabilities may be avoided
by certain constructions in string theory --- see, \eg ref.~\cite{don}.

However, this result on the necessity of negative tension branes cannot
be completely general as it is straightforward to construct consistent
compactifications in six or higher dimensions which only involve positive
tension branes (as will be discussed in detail below). Of course, there
is no mystery here as a key ingredient in ref.~\cite{gibbons}
was that the analysis was limited to theories in {\it five dimensions!}

The purpose of this paper is to extend the consistency conditions to
brane world scenarios in an arbitrary number of spacetime dimensions.
We present these calculations in section 2, where we also demonstrate
how contributions from, \eg non--vanishing curvature on the internal space
allow consistent compactifications to be constructed with only positive
tension branes.  In section 3, we examine how the ``sum rules'' are satisfied
in detail in some six--dimensional models. In particular, we consider
an interesting warped compactification based on the AdS soliton\cite{soliton}
which realizes the Randall--Sundrum hierarchy mechanism with only positive
tension branes. A very similar brane world scenario was considered earlier by
ref.~\cite{old}. We close in section 4 with a discussion of our results
and some concluding remarks.

\section{Consistency Conditions}

Following the approach of ref.~\cite{gibbons}, we use Einstein's equations to
derive some general formulae for the consistency of brane world models 
with a compact internal space. The full spacetime will be $D$--dimensional.
The metric will have a warped product ansatz: 
\beq
\label{metric}
ds^2=G_{MN}(X)dX^MdX^N
    =g_{mn}(y)dy^mdy^n+W^2(y)g_{\mu\nu}(x)dx^\mu dx^\nu,
\eeq
where $X^M$ denote coordinates on the full $D$--dimensional space,
the $p+1$ coordinates $x^{\mu}$ denote the uncompactified directions
in the spacetime, and the remaining $D-p-1$ coordinates $y^{m}$ specify the
compact internal space. As some examples: for $D=5$, $p=3$ and
$W(y)=e^{-2k|y|}$, the above metric corresponds to that studied by Randall and
Sundrum\cite{RS1,RS2} where two copies of a portion of five--dimensional
anti--de Sitter space (AdS$_5$) are pasted together along three--brane
boundaries. When $p=3$ and $W=1$ the metric is factorizable
as considered in ref.~\cite{arkani}. Note that in the latter case, 
the internal compact space is usually considered to be a $(D-p-1)$--dimensional 
torus. We will allow for some generalizations in section 3.

 In keeping with the standard nomenclature, we will
refer to $x^{\mu}$ as the brane coordinates. However, in the following we
will allow for the possibility that the model includes $q$--branes with
$q>p$ and which are, therefore,
extended in some of the internal space directions
as well. The latter, of course, arises in many interesting string theoretic
models\cite{restring}. For a $D$--dimensional model, one could consider
$q$--branes with $q$ as large as $D-1$, which would then be extended in
all of the spacetime dimensions. In our analysis below, the net effect
of such a space--filling brane would be to modify the cosmological constant,
and so generally we will only discuss $q$--branes with $p\le q\le D-2$.

The components of the Ricci tensor in the full spacetime are related to 
their lower dimensional counterparts by
\beqa
\bk{R}_{\mu\nu}&=&\br{R}_{\mu\nu}-\frac{g_{\mu\nu}}{p+1}\frac{1}{W^{p-1}}
\nabla^2 W^{p+1},\label{brRicci}\\
\bk{R}_{mn}&=&\inte{R}_{mn}-\frac{p+1}{W}\nabla_m\nabla_n W,
\label{intRicci}
\eeqa
where $\br{R}_{\mu\nu}$ is the Ricci tensor derived from $g_{\mu\nu}$ 
(independent of the warp factor) and $\inte{R}_{mn}$ is the Ricci tensor 
derived from the internal metric $g_{mn}$. Here $\nabla_{m}$ and $\nabla^{2}$
are respectively the
covariant derivative and the covariant Laplacian with respect to this internal
metric. The three relevant Ricci scalars are denoted
\beq
\bk{R}\equiv\bk{R}_{MN}G^{MN},\quad
\cR\equiv\br{R}_{\mu\nu}g^{\mu\nu},\quad
\ti{\cR}\equiv\inte{R}_{mn}g^{mn}.\label{Rscals}
\eeq
Taking partial traces in eqs.~\reef{brRicci} and \reef{intRicci}
yields
\beqa
\frac{1}{p+1}\left(\cR W^{-2}-\bk{R}^\mu{}_\mu\right)&=&pW^{-2}
\nabla W\cdot\nabla W+W^{-1}\nabla^2W,\label{brTr}\\
\frac{1}{p+1}\left(\ti{\cR}-\bk{R}^m{}_m\right)&=&W^{-1}\nabla^2 W,
\label{intTr}
\eeqa
where we use the notation: $\bk{R}^\mu{}_\mu\equiv W^{-2}g^{\mu\nu}
\bk{R}_{\mu\nu}$ and $\bk{R}^m{}_m\equiv g^{mn}\bk{R}_{mn}$. Therefore
$\bk{R}=\bk{R}^\mu{}_\mu+\bk{R}^m{}_m$. 
Now, consider the following total derivative on the internal space:
\beq
\nabla\cdot(W^\al\nabla W)
=W^{\al+1}\left[\al W^{-2}\nabla W\cdot\nabla W+W^{-1}\nabla^2W\right],
\label{totder}
\eeq
where $\al$ is an arbitrary constant.
Comparing this with the RHS's of eqs.~\reef{brTr} and \reef{intTr}, we find
that we can write the total derivative as
\beq
\nabla\cdot(W^\al\nabla W)
=\frac{W^{\al+1}}{(p+1)p}\left[\al(\cR W^{-2}-\bk{R}^\mu{}_\mu)
+(p-\al)(\ti{\cR}-\bk{R}^m{}_m)\right].\label{protocond}
\eeq

The full $D$--dimensional Einstein equations may be written as
\beq
\bk{R}_{MN}=8\pi \GC \left(\bk{T}_{MN}-\frac{1}{D-2}G_{MN}\bk{T}^P{}_P\right),
\label{bkEin}
\eeq
where $\GC$ is the $D$--dimensional gravitational constant.
Using these, we can write $\bk{R}^\mu{}_\mu$ and $\bk{R}^m{}_m$ in terms of
the stress--energy tensor:
\beqa
\bk{R}^\mu{}_\mu&=&
\frac{8\pi \GC}{D-2}\left((D-p-3)\bk{T}^\mu{}_\mu-(p+1)\bk{T}^m{}_m\right),
\label{brT}\\
\bk{R}^m{}_m&=&\frac{8\pi \GC}{D-2}\left((p-1)\bk{T}^m{}_m-
(D-p-1)\bk{T}^\mu{}_\mu\right),
\label{intT}
\eeqa
where we have used $\bk{T}^M{}_M=\bk{T}^\mu{}_\mu+\bk{T}^m{}_m$ --- that is,
$\bk{T}^\mu{}_\mu\equiv W^{-2}g^{\mu\nu}\bk{T}_{\mu\nu}$, in analogy with the
above. Substituting eqs.~\reef{brT} and \reef{intT} into eq.~\reef{protocond},
the total derivative becomes
\beqa
\nabla\cdot(W^\al\nabla W)
&=&\frac{W^{\al+1}}{(p+1)p}\Bigg\{
\frac{8\pi \GC}{D-2}\bigg(\bk{T}^\mu{}_\mu[(p-2\al)(D-p-1)+2\al]\non\\
& &\qquad\qquad+\bk{T}^m{}_m[2\al-p(p-1)]\bigg)+(p-\al)\ti{\cR}+\al \cR W^{-2}
\Bigg\}\ .
\label{condition1}
\eeqa
If we have a compact internal space, the integral of the LHS
vanishes.\footnote{Some attention should be paid to the smoothness of the
warp factor.} We are then left with
\beqa
&&\oint W^{\al+1} \bigg(\bk{T}^\mu{}_\mu\big[(p-2\al)(D-p-1)+2\al\big]
+\bk{T}^m{}_m\,p(2\al-p+1)\non\\
&&\qquad\qquad+\frac{D-2}{8\pi \GC}\big[(p-\al)\,\ti{\cR}
+\al\, \cR\, W^{-2}\big]\bigg)\ =\ 0,
\label{conduit}
\eeqa
which is a constraint that must be satisfied by any consistent brane world
model. Setting $\al=n-1$, $p=3$ and $D=5$ (for which
$\ti{\cR}=0$) reproduces the consistency conditions derived in
ref.~\cite{gibbons}. Eq.~\reef{conduit} provides a generalization
of their work, which in particular is not limited
to internal spaces of one dimension.
Finally, if the internal space is not compact, eq.~\reef{conduit} may still
provide an interesting consistency condition, as long as care is taken with
the boundary conditions.

We wish to apply this condition to various brane world scenarios to test
their consistency with Einstein's equations. With this in mind, we write an
ansatz for the stress--energy tensor of the form
\beq
\bk{T}_{MN}=-\frac{\Lambda G_{MN}}{8\pi \GC}- 
\sum_i T_q^{(i)} P[G_{MN}]_q^{(i)}\D^{(D-q-1)}(y-y_{i})+\cT_{MN}\ .
\label{genstressans}
\eeq
As well as a bulk cosmological constant,
this describes a collection of branes of various dimensions. The $i^{\mt{th}}$
brane is a $q$--brane (with $q\ge p$) with tension $T_q^{(i)}$ (with units
energy/length$^q$) and transverse
coordinates $y_i$. $P[G_{MN}]_q^{(i)}$ is the pull--back of the spacetime
metric to the worldvolume of the $q$--brane. Any other bulk or worldvolume
matter field
contributions are implicitly encoded in $\cT_{MN}$. In this ansatz, 
$\D^{(D-q-1)}(y-y_i)$ denotes
that covariant combination of delta functions and (geo)metric factors necessary
to position the brane. Typically, this will be a product of terms of the form
$\del(y-y_i)/\sqrt{G_{yy}}$, but a more sophisticated expression may be required
if some of the relevant coordinates are ignorable at the position of the brane
--- see appendix A. Note that we are implicitly assuming that all of the 
branes are extended in all of the $x^\mu$ directions, and, if $q>p$ for
a particular brane, it spans a $(q-p)$--cycle in the internal space.

Given this ansatz, we deduce that
\beqa
\bk{T}^\mu{}_\mu&=&-(p+1)\left[\frac{\Lambda}{8\pi \GC}+\sum_i T_q^{(i)}
\D^{(D-q-1)}(y-y_i)\right]+\cT^\mu{}_\mu\ ,
\label{brstress}\\
\bk{T}^m{}_m&=&-(D-p-1)\frac{\Lambda}{8\pi \GC}-\sum_i (q-p)\,T_q^{(i)} 
\D^{(D-q-1)}(y-y_i)+\cT^m{}_m\ .\label{intstress}
\eeqa
The consistency condition \reef{conduit} may now be written as
\beqa
&&\oint W^{\al+1} \bigg(\al\, \cR\,W^{-2}+(p-\al)\,\ti{\cR}
-\left[\gamma+(D-p-1)\,\ti{\gamma}\right]\Lambda
\label{condition2}\\
&&\qquad\qquad-8\pi \GC\left[\displaystyle \sum_i \left(\gamma+(q-p)\,
\ti{\gamma}\right)T_q^{(i)}\D^{(D-q-1)}(y-y_i)
-\frac{\gamma}{p+1}\,\cT^\mu{}_\mu-\ti{\gamma}\,\cT^m{}_m\right]
\bigg)\ =\ 0\ ,
\non
\eeqa
where we have introduced the following constants:
\beq
\gamma=\frac{p+1}{D-2}\left[(p-2\al)(D-p-1)+2\al\right]\ ,\qquad
\ti{\gamma}=\frac{p(2\al-p+1)}{D-2}\ .
\label{gammas}
\eeq

Dimensional parameters aside, eq.~\reef{condition2} gives a one parameter
($\al$) family of consistency conditions relating
the geometry of the brane world to its stress--energy content. As such, this is
merely a convenient re--expression of certain components of Einstein's
equations, with a
stress--energy tensor of the form \reef{genstressans}. These general results
are not particularly transparent and so to get a better insight into what
these sum rules are telling us, we will now specialize to the
phenomenologically interesting case $p=3$. Focusing on the choice
$\al=-1$ also simplifies the expressions because the warp factor is removed
from all of the terms except that involving $\cR$, the Ricci scalar
for the noncompact metric $g_{\mu\nu}$. With these choices,
eq.~\reef{condition2} reduces to
\beqa
&&\oint \Bigg(-\cR W^{-2}+4\ti{\cR}-\frac{8(D-5)}{D-2}\Lambda
+\frac{8\pi \GC}{D-2} \left[(5D-22)\cT^\mu{}_\mu-12\cT^m{}_m\right]\Bigg)
\non\\
&&\qquad\qquad\quad
=\frac{32\pi \GC}{D-2}\displaystyle \sum_i (5D-13-3q)L_i\,T_q^{(i)}\ ,
\label{p3,al-1}
\eeqa
where $L_i$ is the area of the $(q-p)$--cycle in the internal space
spanned by the $i^{\mt{th}}$ brane. If $q=p$ (\ie the brane is not extended
in the internal space), then $L_i=1$.

Let us first consider the case $D=5$. The above constraint then simplifies, 
since the coefficient of the $\Lambda$ contribution vanishes and $\ti{\cR}=0$
since there is a single internal direction. If we set aside the additional
contributions of matter fields (\ie set $\cT^\mu{}_\mu=0=\cT^m{}_m$), the
constraint becomes
\beq
-\cR \oint  W^{-2}=32\pi G_{5} \sum_{i} T_3^{(i)}\ .
\label{d5cons}
\eeq
Our result here essentially reproduces that given in ref.~\cite{gibbons} ---
compare to their eq.~(2.26). In particular, if the curvature
on the branes is positive or vanishes, we have $\sum_{i} T_3^{(i)}\le 0$
and so we must include some number of negative tension branes for a consistent
model.

Note that in general $\cR=\cR(x)$ is independent of the internal
coordinates $y^n$, hence it appears outside of the integration in
eq.~\reef{d5cons}. However, with the restrictions imposed above
(\ie $\cT_{MN}=0$), it must
further be true that in fact $\cR$ is a fixed constant. The same will be
true in all of the examples considered in the following.

However, let us consider the constraint with $D=6$ but again no matter fields
for simplicity. With these choices, eq.~\reef{p3,al-1} becomes
\beq
\oint \left(-\cR W^{-2}+4\ti{\cR}-2\Lambda\right)
=8\pi G_{6} \displaystyle \sum_i (17-3q)L_i\,T_q^{(i)}\ .
\label{d6cons}
\eeq
Note that on the RHS, there are contributions coming from three-- and
four--branes, both with positive coefficients. On the LHS, however,
we also have contributions coming from the cosmological constant and
the curvature of the two--dimensional internal space. Certainly, these
contributions afford us much more leeway in constructing consistent
brane world models, even when no matter fields are present.
For instance, a positive $\ti{\cR}$ and negative $\L$ can produce
an overall positive contribution on the LHS which could then accommodate
the appearance of only positive tensions on the RHS. Similar contributions
from the cosmological constant and internal curvature appear in
eq.~\reef{p3,al-1} for all higher dimensions $D\ge6$.
Hence, the sum rules are obviously much less restrictive when we go
beyond $D=5$.  We shall explore a few examples in the following section.  

\section{$D=6$ Brane World Examples}

As an application of the sum rules \reef{condition2} derived above,
we consider two examples with $D=6$ and $p=3$. Other interesting
examples may be found in refs.~\cite{old,sixpos,nelson,sixpop}. In particular,
refs.~\cite{old,sixpos} provide compactifications involving only positive
tension branes. With the present choice of dimensions, the constants in
eq.~\reef{gammas} become
\beq
\gamma=2(3-\al)\ ,\qquad\ti{\gamma}=\frac{3}{2}(\al-1)\ ,\qquad\gamma+
(D-p-1)\,\ti{\gamma}=3+\al\ .\label{p3,D6gammas}
\eeq
For matter fields, we only consider a gauge field whose field strength is
proportional to the volume form of the two-dimensional internal
space, \ie $F_{mn} = k \epsilon_{mn}$. The traces of the 
corresponding stress--energy tensor are then
\beq
\label{gauget1}
\cT^{\mu}{}_{\mu} = -k^{2}\ ,\qquad \cT^{m}{}_{m} = {k^{2}}/{2}\ .
\eeq 
The general consistency conditions \reef{condition2} then become
\beqa
&&\oint W^{\al+1} \left(\al\, \cR\,W^{-2}+(3-\al)\,\ti{\cR}
-(3+\al)\Lambda-2\pi \GC(9-5\al) k^2\right)
\label{p3,D6intcond}\\
&&\qquad\qquad\quad=4\pi G_{6} \left(4(3-\al)\displaystyle \sum_i T_3^{(i)}
W_i^{\al+1}
+(9-\al)\displaystyle\sum_i T_4^{(i)}\oint_iW^{\al+1} \right)
\non
\eeqa
where $W_i=W(y=y_i)$ and $\oint_iW^{\al+1}$ denotes an integral over the
one--cycle spanned by the $i^{\mt{th}}$ four--brane in the internal space.

Considering again the choice $\al=-1$, we have a slight generalization
of eq.~\reef{d6cons}:
\beq
\al=-1:\qquad- \cR \oint W^{-2}+4 \oint \ti{\cR}
-2\Lambda V_2-28\pi\GC k^2V_2=8\pi G_{6}\Big(8\displaystyle \sum_i T_3^{(i)}
+5\displaystyle\sum_i L_i T_4^{(i)}\Big)\ .
\label{-1}
\eeq
Here we have introduced $V_2$ to denote the volume of the internal space,
which appears in the contributions from the cosmological constant and the
gauge field. In this particular equation, the integral over the internal
curvature deserves special attention because it yields a topological
invariant, the Euler character:
\beq
\label{GB1}
\chi=\frac{1}{4\pi} \oint \ti{\cR}\ .
\eeq
Hence, this contribution yields a simple global constant which characterizes
the six--dimensional model of interest.

Returning to the general expression \reef{p3,D6intcond}, we see that
certain choices of $\al$ will cause one or more of the contributions
to vanish. We consider explicitly the following cases:
\beqa
\al=-3:&\quad&\oint W^{-2}\left(- \cR W^{-2}+2\ti{\cR}-
16\pi G_{6} k^2\right)
\non\\
&&\qquad\qquad
=16\pi \GC\Big(2\displaystyle \sum_iT_3^{(i)} W_i^{-2}+
\displaystyle\sum_i T_4^{(i)}\oint_iW^{-2}\Big)\ ,
\label{-3}\\
\al=0:&\quad&\oint W\left(\ti{\cR}-\Lambda-
6\pi G_{6} k^2\right)=4\pi \GC\Big(4\displaystyle \sum_i T_3^{(i)}W_i+
3\displaystyle\sum_i T_4^{(i)}\oint_iW \Big)\ ,
\label{0}\\
\al=\frac{9}{5}:&\quad&\oint W^{14/5}\left(3 \cR W^{-2}+2\ti{\cR}-
8\Lambda\right)
\non\\
&&\qquad\qquad
=16\pi G_{6}\Big(2\displaystyle \sum_i T_3^{(i)}W_i^{14/5}+
3\displaystyle\sum_i T_4^{(i)}\oint_iW^{14/5} \Big)\ ,
\label{9/5}\\
\al=3:&\quad&\oint W^4\left(\cR W^{-2}-2\Lambda+4\pi \GC k^2\right)=
8\pi G_{6}\displaystyle\sum_i T_4^{(i)}\oint_iW^{4}\ ,
\label{3}\\
\al=9:&\quad&\oint W^{10}\left(3\cR W^{-2}-2\ti{\cR}-4\Lambda+24\pi \GC k^2
\right)=
-32\pi G_{6}\displaystyle\sum_i T_3^{(i)}W_i^{10}\ .
\label{9}
\eeqa
We will comment on these expressions for the following two examples.

\subsection{Non--warped example}

We consider first the example of non--warped compactifications,
\ie compactifications with a factorizable spacetime manifold, as arose
in the original discussion of large extra dimensions\cite{arkani}.
That is, we set $W=1$ everywhere. Of course, this greatly simplifies the
consistency conditions above.

One of the interesting aspects of six--dimensional models is that three--branes
are co--dimension two objects. Therefore, the effect of a relativistic
three--brane on the spacetime geometry is to induce an angular deficit
in the transverse space, in analogy to the effect of a cosmic string in four
dimensions\cite{cosmic}. That is, locally the geometry of the
internal space is a cone with the three--brane located at the tip\cite{deser}.
Einstein's equations relate the local deficit angle to the tension of the
three--brane by
\beq
\delta_i = 8\pi G_{6} T_3^{(i)}.\label{tangle}
\eeq  

As a result of this simple geometric effect, it is straightforward to
produce consistent compactifications which involve only flat three--branes.
If we include only three--branes and set $\L=0=k=\cR$ (as well as $W=1$),
then all of the above consistency conditions yield a single nontrivial
constraint,
\beq
\chi = 4 G_{6} \sum_i\;T_3^{(i)},
\label{conone}
\eeq
where we have used eq.~\reef{GB1}. If the internal space has a spherical
topology, so that
$\chi=2$, we can construct a brane world model with only positive tension
three--branes. Upon using eq.~\reef{tangle}, this consistency condition becomes
\beq
\sum_i \delta_i = 4\pi,
\eeq
which was previously noted in this context in ref.~\cite{drum}.
If instead we consider a torus ($\chi=0$), as is usually
considered in these scenarios\cite{arkani}, consistency would demand that
we introduce a number of negative tension three--branes. The same would
be true for internal spaces of higher genus, as arise, \eg in the
compactifications considered in ref.~\cite{hyper}. Note that in all
the cases discussed here, the spacetime is locally completely flat.
The three--branes only introduce delta--function curvature distributions
in the transverse space which produce a compact internal manifold.

Note that eq.~\reef{conone} puts no constraints on the number of branes that
might appear in, \eg the two-sphere compactification. Geometrically, however,
there is a lower bound. If we want a smooth internal space with a finite
volume, we would need at least three branes or curvature sources at distinct
positions. Of course, this is essential if the model is to produce interesting
phenomenology, as the four--dimensional Newton's constant $G_{4}$
is related to that in six dimensions by\cite{arkani}
\beq
\label{gravi}
G_{4} = {G_6}/{V_{2}}.
\eeq   

Let us now generalize the discussion of these factorizable scenarios
by including the contributions of the cosmological constant and the
gauge field flux on the internal space. We still only include flat
three--branes with $\cR=0$. In this case, the previous analysis imposes
two nontrivial constraints. The first comes most directly from eq.~\reef{3}
which yields
\beq
\Lambda=2\pi G_{6} k^2 >0.
\label{lamchop}
\eeq
Essentially, this constraint on the internal flux and $\Lambda$ ensures that
Einstein's equations are still satisfied everywhere with a Ricci--flat
brane metric $g_{\mu\nu}$. The remaining constraint generalizes
eq.~\reef{conone} to
\beq
\chi -2 G_{6} V_2k^2= 4 \GC \sum_i\;T_3^{(i)}.
\label{contwo}
\eeq
Hence, in some sense, the introduction of these extra parameters
makes harder the construction of consistent compactifications with only
positive tension branes. Of course, consistent models are still possible
with a compactification on a two--sphere. Ref.~\cite{leblond} provides
an explicit realization as follows: 
After a compactification on a round two--sphere where eq.~\reef{lamchop}
is satisfied, we cut out a wedge along two meridians. Pasting these two edges
introduces angular deficits representing
three--branes with equal tensions, at the north and south poles. Of course,
more elaborate constructions involving more three--branes would also be
possible.

\subsection{Brane worlds from the AdS soliton}

In this section, we consider a less trivial six--dimensional configuration in
order to demonstrate the use of the consistency conditions. 
We begin by describing, in some detail, the geometry of the AdS
soliton\cite{soliton}. This is an asymptotically locally AdS solution of 
Einstein's equations with a negative cosmological constant. We demonstrate
how, by applying a cutting--and--pasting procedure analogous to that used
in the Randall--Sundrum scenario\cite{RS1,RS2} (see discussion in
ref.~\cite{roberto}), the AdS soliton yields a brane world model with
a two--dimensional internal space of spherical topology. A
similar model was considered earlier in ref.~\cite{old} from a
different point of view.
Finally we consider below the consistency conditions as they apply to
the resulting solution.

The AdS soliton can be constructed by doubly 
analytically continuing the metric for a planar AdS--Schwarzschild black hole
in $D$ dimensions. Here we focus our attention on the case $D=6$ which will
lend itself to producing a (3+1)--dimensional brane world.
In horospheric coordinates, the line element is
\beq
ds^2=\frac{r^2}{L^2}(\eta_{\mu\nu}dx^\mu dx^\nu+f^2(r)d\tau^2)
+\frac{L^2}{r^2}\frac{dr^2}{f^2(r)},\label{sol}
\eeq
where $f^2(r)=1-\w^5/r^5$ and $\eta_{\mu\nu}$ is the four--dimensional Minkowski 
metric. $L$ is related to the cosmological constant by: $\Lambda=-10/L^2$.
By comparison with eq.~\reef{metric}, the warp factor is $W(r)=r/L$.
The $\tau$ coordinate, which will become a part of an internal space,
will be periodically identified with period
\beq
\ell=\frac{2 L^2}{5\w}\left(2\pi-\delta\right).\label{singper}
\eeq
As $f(r)$ vanishes at $r=\w$, this circle shrinks to zero size and the
geometry closes off there. Hence, in the basic AdS soliton, we only consider
the radial coordinate for the range $r\ge \w$. The parametrization was
chosen in eq.~\reef{singper} such that if $\delta=0$, the geometry is actually
smooth at $r=\w$. Otherwise, the spacetime exhibits a conical singularity in
the $r,\tau$ space, with
an angular deficit $\delta$ at $r=w$. In the current context, we interpret
this geometry as arising from a three--brane positioned at the tip of the
cone with tension
\beq
\delta = 8\pi G_6 T_3\ ,\label{tangledd}
\eeq 
in analogy to the discussion in the previous subsection.
 
To construct a brane world geometry with a compact internal space, we can
truncate the AdS soliton at some value of the radius, say $r=R$, and paste two
copies of the inner region along the cutting surface. The $r$ and $\tau$
coordinates then form an internal space with spherical 
topology. In the following, we actually consider a slightly more elaborate
configuration by letting the two regions which are being pasted together
have different parameters, $\w$, $R$ and $\delta$.\footnote{The model
of ref.~\cite{old} corresponds to the case where a $Z_2$ symmetry is
imposed at the boundary. A further extension
of our construction would be to allow $L$ or the cosmological constant have
different values on either side of the interface, as might happen for
certain types of domain walls.} To paste together the two geometries
smoothly requires us to match the metrics induced on the interface by the 
different embeddings. Consider joining one copy with $\w=\w_1$, $R=R_1$ and
$\delta=\delta_1$  to another with $\w=\w_2$, $R=R_2$ and $\delta=\delta_2$.
At the interface the $x^\mu$ coordinates in the second copy can be Lorentz
transformed so that they are aligned with those in the first copy, but in
general the warp factors in the two geometries will differ at the
interface. Hence, matching the metrics will require that we introduce a scaling
$x_2^\mu=\beta x_1^\mu$ where
\beq
\beta={W(r_1=R_1)\over W(r_2=R_2)}=\frac{R_1}{R_2}\ .
\label{xmatch}
\eeq
Similarly, the $\tau$ coordinates on either side should be scaled to match
across the interface: $\tau_2=\ell_2\tau_1/\ell_1$. Further,
the metric parameters must be constrained in order that the proper period of
the circle direction is the same in both of the geometries:
\beq
\left(\frac{R_1L_1(2\pi-\delta_1)}{\w_1}\right)^2\left(1-\frac{\w_1^5}{R_1^5}
\right)=\left(\frac{R_2L_2(2\pi-\delta_2)}{\w_2}\right)^2\left(1-
\frac{\w_2^5}{R_2^5}\right)\ .
\label{tmatch}
\eeq

Imposing these conditions will ensure continuity of the metric at the interface,
but in general, it will not be differentiable. The discontinuity in the
extrinsic curvature across the interface is interpreted as resulting from a
delta--function source of stress--energy distributed over this
hypersurface\cite{israel} --- see also ref.~\cite{mtw}. The surface
stress--energy tensor $S_{AB}$ may be calculated as
\beq
8\pi G_{6} S_{AB}=(K^{(2)}+K^{(1)})_{AB}-G_{AB}\,(K^{(2)}+K^{(1)})^C{}_C\ ,
\label{hypEin1}
\eeq
where $K^{(i)}_{AB}=n_{i}\pf_{r_i}G_{AB}/2$ are the extrinsic curvatures of the 
interface due to its embedding in each of the geometries. Note that the
latter formula for the extrinsic curvature applies because of the diagonal
form of the metric. The normalization factors in this expression are given by
\beq
n_{i}=\frac{R_i}{L}\sqrt{1-\w^5/R_1^5}\ .
\eeq
The surface stress--energy is then found to have components
\beqa
S_{\mu\nu}&=&-\frac{\eta_{\mu\nu}}{16\pi G_6 L^3}R_1^2\left\{
\left(1-\frac{\w_1^5}{R_1^5}\right)^{-1/2}
\left[8-3\frac{\w_1^5}{R_1^5}\right]\right.\non\\
& &\qquad\qquad\qquad\qquad\left.+\left(1-\frac{\w_2^5}{R_2^5}\right)^{-1/2}
\left[8-3\frac{w_2^5}{R_2^5}\right]\right\},\label{Smn}\\
S_{\tau\tau}&=&-\frac{1}{2\pi G_6 L^3}
\left\{\left(1-\frac{w_1^5}{R_1^5}\right)^{3/2}R_1^2-
\left(1-\frac{\w_2^5}{R_2^5}
\right)^{3/2}\left[R_2\frac{\ell_2}{\ell_1}\right]^2\right\},
\label{Stt}
\eeqa
where use has been made of the matching conditions \reef{xmatch} and
\reef{tmatch}.

To interpret this surface stress--energy as arising from a(n infinitely thin)
relativistic four--brane, it should have the form $S_{AB}=-T_4G_{AB}$.
Unfortunately, eqs.~\reef{Smn} and \reef{Stt} do not accommodate this simple
interpretation. However, the problem is solved naturally by assuming that
the source is composed of a bound state of a four--brane and a three--brane
delocalized around the circle direction. If we assume that the brane tensions
combine linearly,\footnote{This would result if the four--brane theory
involved a three--form gauge potential with a simple quadratic action:
${\cal L}_4=-T_4\sqrt{G}(1+F^2)$.} the stress--energy would have the form:
\beqa
S_{\mu\nu}&=&\left.-\left(T_{4}+\frac{T_{3}}{L_\tau}\right)G_{\mu\nu}
\right|_{\mt{interface}}\ ,\label{braneT1}\\
S_{\tau\tau}&=&\left.-T_{4}\,G_{\tau\tau}\right|_{\mt{interface}}\ ,
\label{tauT1}
\eeqa
where $L_\tau$ is the proper period of the circle direction, \ie
$L_\tau=\ell_1\,f(R_1) R_1/L$. We are assuming that the three--brane is
extended in the $x^\mu$ directions, but is also delocalised around the
$\tau$ circle. With this ansatz, we find the tensions to be
\beqa
T_{4}&=&\frac{1}{2\pi G_6 L}\left\{\left(1-\frac{\w_1^5}{R_1^5}\right)^{1/2}
+\left(1-\frac{\w_2^5}{R_2^5}\right)^{1/2}\right\}\ ,\label{tauT2}\\
T^{(3)}_{3}&=&\frac{1}{8\pi G_6}\left\{\left(\frac{\w_1}{R_1}\right)^{4}
(2\pi-\delta_1)
+\left(\frac{\w_2}{R_2}\right)^{4}(2\pi-\delta_2)\right\}\ .\label{braneT2}
\eeqa
Although $T_{4}$ displays no explicit dependence upon the deficit angles,
such a  dependence is implicit through the matching condition \reef{tmatch}.
Note that  both these tensions are positive. We have added a superscript
3 to the three-brane tension, as we have already introduced three--branes
at the two conical singularities at $r_i=\w_i$
with tensions proportional to their deficit angles: 
$T_3^{(i)}=\del_i/(8\pi G_6)$ with $i=1,2$.
In this way, we see that the AdS soliton provides a basic building block for
constructing highly tunable brane world spacetimes where, in principle, any of
the three three--branes might be considered as the `visible' one.  

To further study the details of our six--dimensional brane world, it is
convenient to define a new radial coordinate 
\beq
\rho=\left\{\begin{array}{ll} r_1 & 
\quad\textrm{in the first copy of the AdS soliton,} \\
R_1+R_2-r_2 & \quad\textrm{in the second copy of the AdS soliton.}
                \end{array}\right.
\label{newrad}
\eeq
Then the $\tau$ circle closes off at $\rho=\w_1$, the interface lies at
$\rho=R_1$, and the second copy closes off at $\rho=R_1+R_2-\w_2$. In terms of
this coordinate, the warp factor is
\beq
W(\rho)=\frac{\rho}{L}\theta(R_1-\rho)+(R_1+R_2-\rho)\frac{\beta}{L}
\theta(\rho-R_1).
\eeq
The stress--energy tensor as given in eq.~\reef{genstressans} becomes
\beq
\bk{T}_{MN}=-\frac{\Lambda G_{MN}}{8\pi G_6}-\sum_i^3 T_3^{(i)}
P[G_{MN}]_3^{(i)}\D^{(2)}(y-y_i)-T_4P[G_{MN}]_4\D^{(1)}(y-y_4)
\label{solstress}
\eeq
where the $\D$--functions are given by (see appendix A)
\beqa
\D^{(2)}(y-y_1)&=&\del(\rho-\w_1)/\ell_1\ ,\qquad 
\D^{(2)}(y-y_2)=\del(\rho-[R_1+R_2-\w_2])/\ell_2\ ,\non\\
\D^{(2)}(y-y_3)&=&\del(\rho-R_1)/L_\tau\ ,\qquad
\D^{(1)}(y-y_4)=\del(\rho-R_1)/\sqrt{G_{\rho\rho}}\ .\label{Deltas}
\eeqa
Finally, the internal volume is given by
$V_2=\ell_1(R_1-\w_1)+\ell_2(R_2-\w_2)$, and the internal curvature 
can be shown to be
\beqa
\ti{\cR}&=&16\pi \GC\displaystyle \sum_i^3 T_3^{(i)}\D^{(2)}(y-y_i)+4\pi
\GC T_4\D^{(1)}(y-y_4)\non\\
& &-2\left(\frac{\rho^5-6\w_1^5}{\rho^5L^2}\right)\theta(R_1-\rho)
-2\left(\frac{(R_1+R_2-\rho)^5-6\w_2^5}{(R_1+R_2-\rho)^5L^2}\right)\theta(\rho-R_1),\label{ticR}
\eeqa
either by using eq.~\reef{intTr} and subsequent formulae from section two, or
from the intrinsic properties of the internal space. 

We are now in a position to verify the sum rules \reef{-1} and
(\ref{-3}--\ref{9}). The simplest of these is eq.~\reef{-1}, which becomes
\beq
\label{alpha1}
16\pi+ V_{{\ssc 2}} |\L|=32\pi G_6\displaystyle \sum_{i=1}^3 T_3^{{\ssc (i)}}
+20\pi\GC L_{\tau} T_4\ .
\eeq
Given the results above, an
explicit calculation verifies that this constraint, as well as the remaining
consistency conditions are, indeed, satisfied. Of course, this is actually
a consistency check of our different calculations as we began this subsection
by showing how the brane world
based on the AdS soliton satisfies Einstein's equations.

\section{Discussion}

We have extended the brane world consistency conditions derived in
ref.~\cite{gibbons} for five dimensions to theories of an arbitrary spacetime
dimension. Ultimately, these sum rules amount to a clever re--expression of
certain components of Einstein's equations. However, they prove very
powerful in characterizing five--dimensional models, as it was shown
quite generally that a consistent compactification should include negative
tension branes. The sum rules become less restrictive for $D\ge6$, as
illustrated in eq.~\reef{p3,al-1}. The essential new ingredient in higher
dimensions is that the internal space may have nontrivial curvature which
contributes on the LHS of Einstein's equations. Hence, we found that a
model with a positively curved internal space (\ie $\ti\cR>0$) can be
consistently constructed with only positive tension branes. In
section 3.1, the non--warped compactifications on a two--sphere provide
examples where the internal curvature precisely matches the contribution from
the positive tension branes in the consistency conditions.

In eq.~\reef{p3,al-1}, we see that the coefficient of the cosmological
constant happens to vanish in precisely five dimensions but is negative
for $D\ge 6$. Hence in higher dimensions, a negative cosmological constant
is another ingredient which helps in producing consistent compactifications
with only positive tension branes. This played a role in the warped
compactifications of section 3.2, which were based on the AdS soliton.

An interesting feature of the six--dimensional models is that three--branes
are codimension two objects and so only introduce isolated curvature defects
in the internal space. There is a very simple relation between the
three--brane tension and the internal geometry, as illustrated in
eq.~\reef{tangle}. For higher dimensions (\ie $D>6$), the curvature generated
by the self--gravity of the three--branes is no longer localized. In
particular, if a three--brane was allowed to become arbtrarily
thin, it would be surrounded by an event horizon with $r_H^{D-6}
\sim\GC T_3$. Therefore a realistic brane world must introduce
a model in which the three--branes have a thickness larger than this
horizon radius, and the curvature in the internal space
becomes dependent on this model for the internal structure of the branes
--- see, for example, the discussions in ref.~\cite{twist}. Hence, the
sum rules lose much of their power, in that it is much harder to derive
statements that cover a broad class of models. Of course, we can always
consider the sum rule \reef{condition2} with $\alpha=p$, in which the
coefficient of the internal curvature vanishes. With this choice,
the total derivative in eq.~\reef{totder} is precisely that appearing already
in the the components of the Ricci tensor with brane coordinate
indices (\ie $R_{\mu\nu}$), as given in eq.~\reef{brRicci}.
For the phenomenologically interesting case of $p=3$, the corresponding
sum rule is
\beq
\cR\oint W^{2}-{8\over D-2}\Lambda\oint W^4
={32\pi\GC\over D-2}\left[\displaystyle \sum_i (6-D)T_3^{(i)} W_i^4
+\displaystyle \sum_{i,q>3} (q+3-D)T_q^{(i)}\oint_i W^4\right]\ ,
\label{condition3}
\eeq
where we have not included any matter field contributions. For $D>6$, this
equation tells us that if
we wanted to construct a consistent compactification with only flat
(\ie $\cR=0$) three--branes, we would have to include a {\it positive}
cosmological constant in the theory. Of course, this equation does not
guarantee that such a solution exists but only provides a necessary
condition for consistency. For example, if we applied the same reasoning
to $D=5$, we would conclude that a negative cosmological constant is
necessary. However, examining other consistency conditions, \eg
eq.~\reef{d5cons}, tells us that a consistent solution with only flat,
positive tension three--branes is impossible in five dimensions, independent
of the sign or magnitude of the cosmological constant.
Note that for $D=6$ such a compactification would
not be possible unless $\Lambda=0$, which was the case for our non--warped
examples. For the warped $D=6$ example based on the AdS soliton, we have
a negative
cosmological constant but its contribution in eq.~\reef{condition3} is
balanced by that of the central four--brane on the RHS.

Given that the results of the sum rules are less restrictive in higher
dimensions, one might also gain insight by establishing
inequalities as follows: Multiply the total derivative in eq.~\reef{totder}
by $W^{-\beta}$ and then integrate over the internal manifold. After
integrating by parts, one finds
\beq
\beta\,\oint W^{\alpha-\beta-1}(\nabla W)^2 \ge 0\ ,
\label{inneqq}
\eeq
where the inequality assumes that $\beta$ is positive, and it will only
be saturated if $W$ is a constant. Hence following the same analysis as
in section 2, eq.~\reef{condition2} becomes an inequality with the LHS being 
greater than or equal to zero. The only modification
to the integrand is that the initial factor of $W^{\alpha+1}$ is replaced by
$W^{\alpha+1-\beta}$. Hence we have the freedom to eliminate this term
by choosing $\beta=\alpha+1$ producing an expression where the warp factor
only appears in the first term involving the brane curvature. We
can again eliminate the contribution from the internal curvature by choosing
$\alpha=p$, as above. In this case, with $p=3$ (and $\cT_{MN}=0$),
eq.~\reef{condition3} is replaced by
\beq
\cR\oint W^{-2}-{8\over D-2}\Lambda\,V_{\ssc D-4}
\ge{32\pi\GC\over D-2}\left[\displaystyle \sum_i (6-D)T_3^{(i)} 
+\displaystyle \sum_{i,q>3} (q+3-D)T_q^{(i)}L_i\right]\ .
\label{condition4in}
\eeq
While less precise than the sum rules, inequalities such as these
were sufficient
to  establish certain no--go theorems\cite{berd,juanun} and also played a
role in guiding the construction of ref.~\cite{gkp}.

The warped brane world based on the AdS soliton deserves further comment
as it may be useful in providing a phenomenologically interesting scenario.
First we remark that there are a number of straightforward extensions of
the construction described in section 3.2. First of all, AdS soliton solutions
exist for arbitrary dimensions, and so this construction can be generalized
to produce a warped compactification for arbitrary $p$. Of course, some of
the additional $x^\mu$ would then have to be compactified to produce
a four--dimensional effective theory at low energies. The construction can
also be extended to include a magnetic flux on the compact space by 
beginning with the analogous AdS soliton constructed by an
appropriate analytical continuation of the AdS--Reissner--Nordstr\"{o}m
black hole. Ref.~\cite{old} also considered the extension of
these compactifications such that the brane world has a cosmological metric,
similar to the discussions of ref.~\cite{nem}. 
From our point of view, the essential step is to begin with the standard
AdS--Schwarzschild black hole where the horizon topology is $S^{D-2}\times R$,
rather than $R^{D-1}$ as for the planar black hole. Analytically continuing
and then performing the cut--and--paste construction results in a brane
world where the geometry of the three--branes corresponds to de Sitter space.
Alternatively, anti--de Sitter branes can be produced if one begins with
a ``topological'' black hole where the horizon has negative intrinsic
curvature\cite{topbh}. We should also mention that a portion of the AdS
soliton geometry appeared in a more elaborate cut--and--paste construction in
ref.~\cite{nelson}.

One comment about our warped model is that the low energy theory will
include precisely four--dimensional Einstein gravity. This observation
comes from the fact that the initial AdS soliton metric \reef{sol} can
be generalized to
\beq
ds^2=\frac{r^2}{L^2}\left(g_{\mu\nu}dx^\mu dx^\nu+f^2(r)d\tau^2\right)
+\frac{L^2}{r^2}\frac{dr^2}{f^2(r)}.\label{sol22}
\eeq
This metric, with the same function $f^2(r)=1-\w^5/r^5$, still satisfies
the six--dimensional Einstein's equations
with a negative cosmological constant, as long as the brane metric $g_{\mu\nu}$
is Ricci flat (\ie $\cR_{\mu\nu}(g)=0$). That is, the brane metric must satisfy
the (fully nonlinear)
vacuum Einstein equations in four dimensions. Curving the brane metric
in this way to generalize the original solution is a relatively
general property that applies to warped compactifications which display
four--dimensional Poincar\'e invariance\cite{andrew}. Therefore the warping
does not disturb the emergence of the standard Einstein
gravity in the low energy theory. However, the warp factor does modify
the naive relation between gravitational coupling in four and six dimensions:
\beq
G_6=G_4\,\left({R_1^3-\w_1^3\over 3L^2}\ell_1+
{R_2^3-\w_2^3\over 3L^2}\ell_2\right)\ .
\label{newgg}
\eeq
In comparing this result to eq.~\reef{gravi}, recall that
$V_2=(R_1-\w_1)\ell_1+(R_2-\w_2)\ell_2$ for our warped compactification.

From the point of view of linearized fluctuations, the above discussion
indicates
that four--dimensional gravitons in the brane metric will be a zero mode
of our warped brane world. Using a linearized analysis similar to that of
ref.~\cite{glue}, one finds that there are no other zero mode
fluctuations in the internal metric\cite{stable}. That is, there are no
``scalar'' zero modes that correspond
to varying the size or geometry of the internal space. Hence
once the cosmological constant and the brane tensions are fixed, there is a
unique solution for the internal space. This is interesting because
by going beyond five dimensions, not only have we eliminated the need for
negative tension branes, we have also managed to stabilize the internal
space! This is not a generic feature of higher dimensional
compactifications. There are many moduli in the non--warped example
corresponding to both the volume of the compact space and the relative
position of the three--branes. It would be interesting to better understand
what features of the AdS soliton model were essential in stabilizing the
internal space.

An interesting lesson of the RS I scenario\cite{RS1} is
that one can produce a large hierarchy from the
gravitational redshift between branes, as might
arise in a warped compactification.
The warped compactification considered here
provides another realization of this effect \cite{old}
(without the need to introduce negative tension branes). To make this feature
manifest, we perform the following coordinate transformation,
\beq
y(r) = \frac{2L}{5}\textrm{arccosh}\left(\frac{r}{\w}\right)^{\frac{5}{2}}\ ,
\label{warpsolution}
\eeq
on either side of the interface, which puts the metric \reef{sol} in the form
\beq
ds^{2} = W^2(y) \left(\eta_{\mu\nu}dx^{\mu}dx^{\nu} +
\tanh^2\left({5y\over2L}\right)d\tau^{2}\right) + dy^{2}\ ,
\eeq
where the warp factor is now
\beq
W(y) = \frac{\w}{L} \cosh^{2/5}\left(\frac{5 y}{2L}\right)\ .
\label{wfact}
\eeq
In this new coordinate system, the range $\w \leq r \leq R$,
where $R$ is the position of the bound three/four--brane system, 
corresponds to
\beq
0 \leq y \leq \frac{2L}{5} \textrm{arccosh} \left(\frac{R}{\w}
\right)^{\frac{5}{2}}.
\eeq
If we take the visible brane to be one of the three--branes at either
$r_i=\w_i$ with $i=1,2$, then it is clear from eq.~\reef{wfact} that 
a large redshift is easily generated without introducing any large parameters
in the model. In fact, for $y$ not too large, the warp factor has essentially
the same exponential form as in the Randall--Sundrum scenario, \ie
\beq
W(y) \sim\frac{\w}{L}\exp\left(\frac{y}{L}\right),
\eeq
which is essentially a reflection of the fact that the AdS soliton
is asymptotically locally AdS. In the present construction, this hierarchy
is not an adjustable parameter in the theory, rather it will be fixed
implicitly by the relative tension of the branes (and the value of the
cosmological constant).

One may hope to find a realization of this new warped brane model or
some closely related geometry in string/M--theory. One apparent obstacle
would be that our construction involves both three--branes and four--branes.
In a given type II string theory, the dimensions of the different D--brane species
in a given string theory always differ by two\cite{primer}. However, one
could consider working in the type IIA theory where one finds NS5--branes
as well as D4-- and D6--branes. This may provide a natural framework
to attempt a higher dimensional compactification which provides a close
analog of our model based on the AdS soliton.

\section*{Acknowledgements}
This research was supported in part by NSERC of Canada and Fonds FCAR du
Qu\'ebec. We would like to thank Sean Carroll, Neil Constable, Samir Mathur,
Chung-I Tan and Robert Wald for useful conversations. RCM would like to
thank the ITP in Santa Barbara for hospitality during the final stages
of this work. Research at the ITP was supported in part by the
U.S.~National Science Foundation under Grant No.~PHY99--07949.

\appendix
\section{Delta functions in curved space}

In this appendix we discuss the $\D$--functions used throughout this paper, and show how they can be
calculated. The idea is simple; we need a covariant form for the delta--function so as to maintain its
familiar properties whenever we are working in a curved space. Since this is an issue of coordinate invariance,
the need to modify the $\Delta$--function prescription also occurs in flat space. It is convenient to consider
this case and then make the generalisation to curved space. We will follow the treatment in ref.~\cite{hassani}.

In Cartesian coordinates in flat, two--dimensional space, the definition of the delta--function is
\beq
\int\df x\df y\,F(x,y)\delta(x-x_0)\delta(y-y_0)=F(x_0, y_0),
\eeq
where $F(x,y)$ is some function defined on the plane. If we transform to polar coordinates, define
$H(r,\theta)=F(x(r,\theta), y(r,\theta))$,  and assume
that neither $r$ nor $\theta$ is ignorable at the point $P=(x_0,y_0)=(r_0,\theta_0)$, \ie that the
coordinate transformation is invertible at this point, we get
\beq
\int\df r\df \theta\,r H(r,\theta) \D(r-r_0)\D(\theta-\theta_0)=H(r_0,\theta_0),
\label{deltatrans}
\eeq
which suggests that the $\D$--functions should take the form
\beq
\D(r-r_0)\D(\theta-\theta_0)=\frac{1}{r}\delta(r-r_0)\delta(\theta-\theta_0).
\eeq
In other words, the correct prescription for the $\D$--function should involve a term that cancels the 
coefficients of the metric appearing the measure. Making the obvious generalisation to curved space gives
\beq
\D(\xi-\xi_0)=\frac{1}{\sqrt{G_{\xi\xi}}}\delta(\xi-\xi_0),\label{nignorable}
\eeq
where $G_{\xi\xi}$ is the $\xi$ coefficient of the spacetime metric (assumed diagonal).

The situation is a little more involved if one or more of the coordinates is ignorable at the point $P$.
Consider the case when $P$ is the origin of the flat $(r,\theta)$--plane, meaning that $r_0=x_0=y_0=0$ and 
$\theta_0$ is ignorable. Then $H$ can only be a function of $r$, and eq.~\reef{deltatrans} becomes
\beqa
\int\df r\df \theta\,r H(r) \delta(x)\delta(y)&=&\int\df r\, H(r)\int\df \theta\,r \delta(x)\delta(y)=H(0)\non\\
&\Rightarrow&\int\df \theta\,r \delta(x)\delta(y)=\delta(r).
\eeqa
It follows from this equation that $\delta(x)\delta(y)$ cannot be a function of $\theta$, and we are free to
write
\beq
\delta(x)\delta(y)=\D(r-r_0)\D(\theta-\theta_0)=\frac{\delta(r)}{\int\df \theta\,r}=\frac{\delta(r)}{2\pi r}.
\eeq
In this case we must cancel not only the metric factor in the measure, but also the integral over the ignorable
coordinate. We adopt the same prescription when dealing with ignorable coordinates in curved space. That is,
\beq
\D(\xi^1-\xi^1_0)\cdots\D(\xi^N-\xi^N_0)=
\frac{\delta(\xi^1-\xi^1_0)\cdots\delta(\xi^n-\xi^n_0)}
       {\int\df\xi^{n+1}\cdots\df\xi^N\sqrt{G_{\xi^1\xi^1}\cdots G_{\xi^N\xi^N}}},\label{ignorable}
\eeq
where $\{\xi^{n+1},\ldots,\xi^N\}$ are ignorable. 

This argument can no doubt be generalised to the case of a non--diagonal metric
and put on a firmer
mathematical footing by considering the general transformation properties of the
$\Delta$--function 
(for instance, in a $D$--dimensional space, the full $D$--dimensional delta--function must transform
as a relative tensor of weight $-1$ to ensure that $\int\df^{\mt{D}}\xi \,\delta^{(\mt{D})}(\xi)=1$ is a coordinate
invariant). However, the heuristic discussion given above describes the basic idea and is sufficient for our
purposes. 

We conclude with a derivation of the $\D$--functions in eq.~\reef{Deltas}, where the metric is that of the
AdS soliton, \reef{sol}. Note that we will use the $\rho$ coordinate, defined in eq.~\reef{newrad}, and that
$G_{\rho\rho}G_{\tau\tau}=1$. The radial positions of the two conical singularities, $\rho=w$ and 
$\rho=R_1+R_2-w$, are such that $\tau$ is ignorable. Therefore, we use eq.~\reef{ignorable}:
\beqa
\D^{(2)}(y-y_3^{(1)})&=&\frac{\delta(\rho-w)}{\int\df \tau\sqrt{G_{\rho\rho}G_{\tau\tau}}}
=\frac{1}{l_1}\delta(\rho-w),\non\\
\D^{(2)}(y-y_3^{(2)})&=&\frac{\delta(\rho-[R_1+R_2-w])}{\int\df \tau\sqrt{G_{\rho\rho}G_{\tau\tau}}}
=\frac{1}{l_2}\delta(\rho-[R_1+R_2-w]).
\eeqa
The position of the three--brane on the interface is such that neither $\rho$ nor $\tau$ is ignorable. Hence,
using eq.~\reef{nignorable}, we find
\beq
\D^{(2)}(y-y_3^{(3)})=\frac{\delta(\rho-R_1)\delta(\tau-\tau_0)}{\sqrt{G_{\rho\rho}G_{\tau\tau}}}
=\delta(\rho-R_1)\delta(\tau-\tau_0).
\eeq
The four--brane is in a similar position. It is localised only in $\rho$, which coordinate is never ignorable.
Using eq.~\reef{nignorable} again gives
\beq
\D(y-y_4)=\frac{\delta(\rho-R_1)}{\sqrt{G_{\rho\rho}}}. 
\eeq


\end{document}